\documentclass[showpacs,preprintnumbers,eqsecnum,twocolumn,prd,floatfix,nofootinbib]{revtex4-1}
\usepackage{amssymb}
\usepackage{bm}
\usepackage{epsf}
\usepackage{graphicx}

\newcommand{\be}{\begin{equation}}
\newcommand{\ee}{\end{equation}}
\newcommand{\bea}{\begin{eqnarray}}
\newcommand{\eea}{\end{eqnarray}}
\newcommand\ptl{\partial}

\begin{document}

\title{Quantum mechanics allows setting initial conditions at a cosmological
singularity: Gowdy model example. }

\author{S.L. Cherkas}
\email{cherkas@inp.bsu.by}
 \affiliation{Institute for Nuclear Problems, Bobruiskaya
11, Minsk 220030, Belarus}
\author{V.L. Kalashnikov}
\email{v.kalashnikov@aston.ac.uk}
 \affiliation{Aston Institute of Photonic Technologies, Aston University, Birmingham B4 7ET, UK}
\begin{abstract}
It is shown that the initial conditions in the quasi-Heisenberg
quantization scheme can be set at the initial cosmological
singularity per se. This possibility is provided by finiteness of
some quantities, namely momentums of the dynamical variables, at a
singularity, in spite of infinity of the dynamical variables
themselves. The uncertainty principle allows avoiding a necessity
to set values of the dynamical variables at singularity, as a wave
packet can be expressed through the finite momentums. Influence of
the initial condition set in the singularity in such a way to a
number of gravitons under a vacuum state, arising during later
evolution, is investigated. It is shown that, even choosing of
some special state at the singularity minimizing late time
expansion rate, some amount of gravitons still appear in the late
time evolution.
\end{abstract}

 \maketitle

\section{Introduction}

One of the problems of the relativistic cosmology is the
formulation of the initial conditions for the universe evolution.
A lot have been done in this direction concerning quantum fields
at the classical
 uniform background \cite{Birrell1982}, in particular in
describing the origin of primordial inhomogeneities
\cite{Mukhanov2005,Linde1990} giving the initial conditions on the
last scattering surface for the cosmic microwave background
radiation (see \cite{Dodelson2003} and references given herein).

 The modern
description of the uniform background itself includes the
inflation paradigm \cite{Starobinsky1980,Guth1981,Liddle} which
besides the  description of the   density perturbation values,
successfully solves the problems of horizon and flatness. In
describing the earlier stage of evolution, one encounters  the
problem of the initial conditions again. The well-known Penrouse
theorem \cite{Penrouse1965,Geroch1968,Hawking1970} states that
under quite general conditions, the initial point of the evolution
should be singular.

One of the conditions of the Penrouse  theorem is the energy
condition, which is violated during inflation \cite{Borde1997},
but geodesics remains past uncomplete in this case also
\cite{Borde2003}. The incompleteness of geodesics tells us that
 there is a moment in the past (i.e. singularity) beyond which one cannot
  move in past
 direction. It seems natural to set initial conditions at this
 last point of the backward evolution (initial point of future evolution).
 This seems quite impossible,  at first sight, because the dynamical quantities such as amplitudes of the
 matter fields and scale factor logarithm, turns to infinity at
 the singularity.

It is considered that near the  singularity, at the Planck epoch,
quantum effects are crucial. Thus, the problem of the initial
conditions and the singularity should be considered at the quantum
level
\cite{Hartle1983,Vilenkin1988,Bojowald2001,Bojowald2003,Kiefer2009},
although one could  attempt to avoid singularity at a classical
level \cite{Minkevich2006,Santos2015}.

In relation with the singularity problem we need to discuss some
ways of gravity quantization. According to
\cite{Ashtekar2008,Bojowald2011,Bojowald2012,Husain2004,Tarrio2013},
the loop quantum gravity removes the singularity completely,
including different types of future singularities, such as Big
Rip. The absence of singularities  in loop quantum gravity
originates from the fact that the volume operator (and
consequently the universe scale factor) has a discrete spectrum
bounded below. However, there remains a problem, how to connect
this discrete spectrum
 with the time evolution of the universe, canonical
gravity quantization and, about self-consistency of the loop
quantum gravity itself. Work in this directions is in progress
\cite{Ashtekar2015}.

The canonical quantization of general relativity (GR) leads to the
Wheeler-DeWitt (WDW) equation \cite{DeWitt1967,Wheeler1968}, which
is the analog of the Schr\"{o}dinger equation of the ordinary
quantum mechanics. However, the equation does not contain a time
variable explicitly, so one has to interpret the wave function of
the universe in some way. For instance, one could interpret the
scale factor as time the variable; though, it could not be
considered as a complete solution of the problem because one needs
to describe the evolution of dynamical variables including the
scale factor in time, explicitly.
 For instance, in Ref.  \cite{Tarrio2013}  the
effective Hamiltonian have been deduced by corrections with the
loop quantum gravity effects, and then it was investigated
classically (i.e., to describe time evolution, the authors of Ref.
\cite{Tarrio2013} return to classics).

\begin{figure}[h]
  \includegraphics[width=9cm]{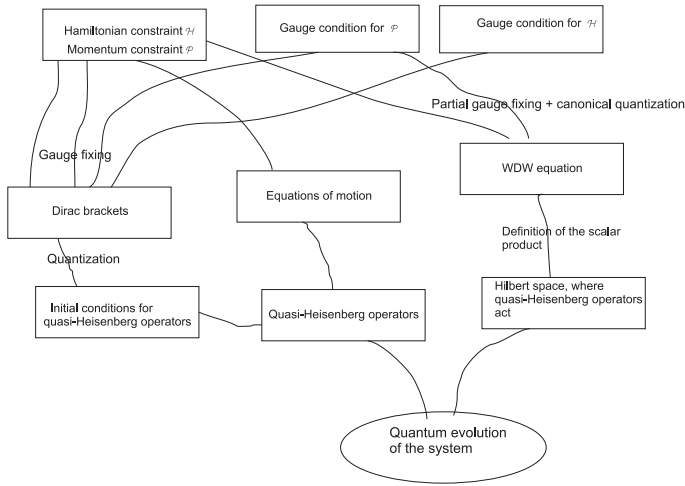}\\
  \caption{Quasi-Heisenberg quntization scheme}\label{H}
\end{figure}

It seems a more fundamental to consider the problem of singularity
and the initial conditions in a quantization scheme involving the
evolution in time explicitly. Such a scheme was suggested for
mini- and midi- superspace models
\cite{Cherkas2006,Cherkas2012,Cherkas2015}. In ordinary quantum
mechanics, Schr\"{o}dinger and Heisenberg pictures are equivalent.
In quantum gravity, a canonically quantized Hamiltonian of the GR
cannot serve for building the Heisenberg picture, that is, the
conventional Heisenberg picture does not exist. Nevertheless, one
can quantize the equations of motion straightforwardly, that is,
quasi-Heisenberg picture exists (Fig. \ref{H}).

In the quantization scheme of Ref.
\cite{Cherkas2006,Cherkas2012,Cherkas2015} quasi-Heisenberg
operators satisfy the commutation relations obtained from the
system of constraints and gauge conditions with the help of the
Dirac brackets at the initial moment of time. Then it is allowed
quasi-Heisenberg operators to evolve according to the equations of
motion. This evolution implicitly determines time-dependent gauge
fixing, defined  explicitly prior to quantization only at initial
moment of time.

It should be noted that the Heisenberg picture for gravity
quantization  using anticommutative ghost variables was discussed
in Ref. \cite{Vereshkov2013}.  The Schr\"{o}dinger picture using
anticommutative ghost variables  has also been developed
\cite{Shestakova1999,Savch,Savch1}. It would be instructive to
compare these approaches with one another and with the
quasi-Heisenberg picture at an example of some a simple
minisuperspace model, but this has not be done yet.

The aim of the present work is to consider more closely the
setting of the initial conditions for the quasi-Heisenberg
operators in connection with the singularity problem.

 Though the initial
singularity remains, the situation differs substantially from the
classic one. It appears, that one may set the initial conditions
at the singularity directly. It will be demonstrated by the
example of the Gowdy model described in section \ref{sect2} of the
paper. This model admits the analytical solution within the whole
time domain and have been used for singularity investigation
\cite{Berger1982,Hussein1987}. Also, this model allows choosing
the out-vacuum state, as the gravitational waves evolve against a
classical background\footnote{ In the general case
quasi-Heisenberg picture admits
 quantum background. }.

Because the existence or nonexistence of the singularity turns out
to be related to the problem of the regularization of the vacuum
energy \cite{Berger1982,Hussein1987}, the issue of vacuum energy
is briefly discussed in section \ref{sect3}, where the evolution
of the system in a vacuum state is considered and then compared
with the evolution in the state given by the wave packet used in
section \ref{sect2}.

\section{Quasi-Heisenberg
quantization of the Gowdy Model}\label{sect2}

The polarized ${\bf T}^3$ Gowdy model corresponds to an
anisotropic universe, where the gravitational waves travel
unidirectionally. Let us take a metric in the form of
\be
ds^2=e^{\tau-\lambda}(d\eta^2-d x^2)-e^{2 \tau+2\sqrt 3
V}dy^2-e^{2 \tau-2\sqrt 3 V}dz^2,
\label{our}
\ee
where the coordinates $\eta,x,y,z$ define points of the
Pseudo-Rimanian manifold. Quantities $\tau, \lambda$ and $V$
determine the manifold metric and depend on the variables $ \eta$
and $x$ only, which takes the values at $\{0,\infty\}$ and $\{0,2
\pi\}$ respectively. We treat the coordinate $\eta$ as a
``time''-parameter describing the evolution of a system.

In Eq. (\ref{our}) we use a slightly different gauge  than the
original Gowdy's one:
\[
ds^2=e^{-\lambda+3\tau}dt^2-e^{-\lambda-\tau}d X^2-e^{2
\tau+2\sqrt 3 V}dy^2-e^{2 \tau-2\sqrt 3 V}dz^2,
\]
where $d t=e^{-\tau} d\eta$ and $dX=e^{\tau} d x$. The motivation
is that in the gauge given by (\ref{our}), the equations of motion
contain a difference of the potential and kinetic energies of
field oscillators. In the absence of evolution, this quantity is
zero by virtue of the virial theorem. When the system evolves, the
virial theorem is violated \cite{Anishchenko2008}.  As was shown
earlier, the difference of the potential and kinetic energies
provides a value of the universe acceleration parameter for the
Friedman universe which is comparable with the observed one
\cite{Cherkas2007}.

The Einstein equations lead to three equations of motion
\bea
V^{\prime\prime}-\ptl_{xx}V+2\tau^\prime
V^\prime-\ptl_{x}V\ptl_{x}\tau=0,~~~~~~~~~~\label{eqmov1}\\
\tau^{\prime\prime}-\ptl_{xx}\tau=2(\ptl_x \tau)^2-2
(\tau^\prime)^2,~~~~~~~~~~~~~\label{eqmov2}\\
\lambda^{\prime\prime}-\ptl_{xx}\lambda=4(\ptl_x \tau)^2-4
(\tau^\prime)^2-6(\ptl_x V)^2+6(V^\prime)^2,~~~~~
\label{eqmov3}
\eea

 and two constraints

\bea
\mathcal   H(\eta,
x)=e^{2\tau}\biggl(\frac{1}{3}(\ptl_x\tau)^2+\frac{1}{2}(\ptl_x
V)^2+\frac{1}{6}\ptl_x\tau\ptl_x\lambda\nonumber\\+\frac{1}{3}
\ptl_{xx}\tau-\frac{1}{3}\tau^{\prime 2}+\frac{1}{2}V^{\prime
2}+\frac{1}{6}\tau^\prime\lambda^\prime \biggr)=0,~~\\
  \mathcal P(\eta, x)=e^{2\tau}\biggl(\frac{1}{6}\ptl_x\lambda\,
\tau^\prime+\ptl_x V V^\prime+\frac{1}{6}\ptl_x \tau
\lambda^\prime~~~~~~~\nonumber\\+\frac{1}{3}\ptl_x\tau^\prime
\biggr)=0,~~
\eea
where prime denotes differentiation over time $\eta$.

Let us discuss the structure of the equations of motion
(\ref{eqmov1})-(\ref{eqmov3}). Eqs. (\ref{eqmov1})-(\ref{eqmov3})
contain a part corresponding to the wave equation. The remaining
parts belong to two different types. The first one is of
$(\tau^\prime V^\prime-\ptl_{x}V\ptl_{x}\tau)$-type. In this case,
we refer to $V$ as a ``field'' variable, whereas $\tau$ plays a
role of the ``background'' against which the field $V$ oscillates.
The equations for the ``background'' variable contain the
difference of the kinetic and potential energies, e.g.,
$(\tau^\prime)^2-(\ptl_x \tau)^2$ or $(V^\prime)^2-(\ptl_x V)^2$.
The situation is analogous to the model representing a string
against a curved background \cite{Cherkas2012}. However, the
equations for the background variable $\tau$ differ from those
considered in Ref. \cite{Cherkas2012}, because Eq. (\ref{eqmov2})
for $\tau$ is isolated, whereas the field variables contribute to
the corresponding equation for the ``background'' in the toy model
\cite{Cherkas2012}. On the other hand, there is another
``background'' variable $\lambda$ here, because the Gowdy model is
anisotropic, and one needs two variables $\tau$ and $\lambda$ to
describe the background. It should be noted that the ``background"
variable $\lambda$ does not influence the oscillations of the
``field" $V$.

In a general case, an inhomogeneous variable $\tau$ has to be
treated as quantum operator with the related algebra. However, the
goal of the present paper is to consider the initial conditions
near singularity. Thus, for simplicity, a particular gauge is
taken where $\tau$ is non-quantum (i.e., ``c''-number valued) and
spatially homogeneous. That results in the solution akin to the
Gowdy one \cite{Gowdy1974,Mizner,Berger}.

It is convenient to expand the dynamical variables into the
Fourier series
\bea
V(\eta,x)=\sum_{k=1} {\mathcal V}_k(\eta) e^{i kx},\nonumber\\
\lambda(\eta,x)=\sum_{k=1} {\Lambda}_k(\eta) e^{i kx},\nonumber\\
\tau(\eta,x)=\sum_{k=1} {T}_k(\eta) e^{i kx} .
\label{cond1}
\eea

\noindent The equation of motion (\ref{eqmov2})  for $\tau$ is
isolated from others. Thus, the spatially uniform initial
conditions for $\tau$ make it spatially independent in the course
of evolution. So one can take the initial conditions
\bea
T_k(0)=\delta_{0,k}{\mathcal T}_0,~~~~
 T^\prime_k(0)=\delta_{0,k}\,e^{-2{\mathcal T}_0}\Pi,
 \label{int}
\eea
\noindent where $\Pi$ and ${\mathcal T}_0$ are some constants. We
shall further refer to ${T}_0(\eta)$ as $\tau(\eta)$.

Advancing in such a way and using the aforementioned gauge, one
comes to the following equations of motion and constraints:

\bea
\tau^{\prime\prime}+2
(\tau^\prime)^2=0,\label{diff1}\\
{\mathcal V}_k^{\prime\prime}+k^2{\mathcal V}_k+2\tau^\prime
{\mathcal V}_k^\prime=0,\label{diff2}\\
\Lambda_0^{\prime\prime}=-4 (\tau^\prime)^2+6\sum_q{\mathcal
V}_q^\prime{\mathcal V}_{-q}^\prime-q^2{\mathcal V}_q{\mathcal
V}_{-q},
\label{diff3}
\eea

\bea
\mathcal H_k=e^{2\tau}\biggl(-\delta_{k,0}\frac{1}{3 }\tau^{\prime
2}+\frac{1}{6}\tau^\prime\Lambda_k^\prime+\frac{1}{2}\sum_q
{\mathcal V}_q^{\prime } {\mathcal V}_{k-q}^{\prime
}\nonumber\\-q(k-q){\mathcal V}_q {\mathcal V}_{k-q} \biggr)=0,~~\label{hk}\\
\mathcal P_k=e^{2\tau}\biggl(\frac{1}{6}i k\Lambda_k\,
\tau^\prime+\sum_q (i q) {\mathcal V}_q {\mathcal
V}_{k-q}^\prime\biggr)=0.~~
\label{pk}
\eea

The equations of motion (\ref{diff1})-(\ref{diff3}) can be
obtained from the Hamiltonian $H=\mathcal H_0$. It should be noted
that $\Lambda_k$ at $k\ne 0$  is completely defined by the
momentum constraint equation (\ref{pk}), namely
\be
\Lambda_k=-\frac{6}{k \tau^\prime}\sum_q q{\mathcal V}_q{\mathcal
V}_{k-q}^\prime,
\ee
which reduces the system to $\tau,\Lambda_0,{\mathcal V}_k$.

One can introduce the momenta
\bea
\pi_k=\frac{\ptl H}{\ptl {\mathcal V}_k^\prime}=e^{2
\tau}{\mathcal V}_{-k}^\prime,~~~
P_\Lambda=\frac{\ptl H}{\ptl \Lambda_0^\prime}=e^{2\tau}\tau^\prime/6,\nonumber\\
P_\tau=-\frac{\ptl H}{\ptl
\tau^\prime}=e^{2\tau}\left(\frac{2}{3}\tau^\prime-\Lambda^\prime_0/6\right),~~
\eea
and rewrite the Hamiltonian  in terms of these momentums
\bea
H=e^{-2\tau}\left(-6P_\Lambda P_\tau+12
P_\Lambda^2+\frac{1}{2}\pi_0^2+\sum_{k\ge 1}\pi_k \pi_{k}^*
\right)\nonumber\\+e^{2\tau}\sum_{k\ge 1}k^2{\mathcal V}_k
{\mathcal V}^*_{k},~~~
\label{hsm}
\eea
where it is taken into account that $\pi_{-k}=\pi_k^*$ and
${\mathcal V}_{-k}={\mathcal V}_k^*$.

The quasi-Heisenberg quantization consists in quantization of the
equations of motion \cite{Cherkas2006,Cherkas2012,Cherkas2015}.
Briefly, this procedure can be described in the following way. The
operator initial conditions for the equations of motion  include
the conditions (\ref{int}) rewritten in terms of $\tau$ and the
remaining conditions
\bea
{\hat {\mathcal V}_k}(0)=\hat v_k,~~~
 \hat {\mathcal V}_k^\prime(0)=e^{-2{\mathcal T}_0}\hat
 p_{-k},~~
\hat  \Lambda_0(0)=L_0,\nonumber\\
 \hat \Lambda_0^\prime(0)=e^{-2{\mathcal T}_0}(24 P_\Lambda(0)-6
 \hat P_\tau(0)),~~~
 \tau(0)={\mathcal T}_0,\nonumber\\
\tau^\prime(0)=6 e^{-2{\mathcal
T}_0}P_\Lambda(0),~~~~~~~~~~~~~~~~~~~~~~~~~~~~~~~~~~
\eea
where ${\mathcal T}_0$ and $L_0$ are some $c$-numbers,

\bea
P_\Lambda(0)=\Pi,~~~~~~~~~~~~~~~~~~~~~~~~~~~~~~~~~~~~~~~~~~~~~~\nonumber\\
\hat P_\tau(0)=\frac{1}{6\Pi}\left(12 \Pi^2+\frac{1}{2}\sum_k \hat
p_k \hat p_{-k} +e^{4{\mathcal T}_0}k^2 \hat v_k \hat
v_{-k}\right),\nonumber
\eea
 and $\Pi$ is the $c$-number as well.  The operators $\hat p_k$ and $\hat
 v_k$ do not depend on time and
satisfy the standard commutation relations $[\hat p_k,\hat
v_{k^\prime}]=-i\delta_{k,{k^\prime}}$, where
$\delta_{k,k^\prime}$ is the Kronneker symbol.  They are initial
values of the time-dependent operators $\hat
 \pi_k(\eta)$ and $\hat
 {\mathcal V}_k(\eta)$.
  One may implement the
above operator commutation relations by the representation $\hat
v_k=v_k$, $\hat p_k=-i\frac{\ptl}{\ptl v_{k}}$, or by the
representation $\hat p_k=p_k$, $\hat v_k=i\frac{\ptl}{\ptl
p_{k}}$. Thus, one have the following commutator algebra at the
initial moment of time $[\hat \pi_k,\hat {\mathcal
V}_{k^\prime}]=-i\delta_{k,{k^\prime}}$, $[\hat P_\tau,\hat
{\mathcal V}_{k}]=-\frac{i\hat \pi_{-k}}{12 \Pi}=-\frac{i\hat
\pi_{k}^+}{12 \Pi}$,  $[\hat P_\tau,\hat \pi_{k}]=\frac{i k^2
e^{4\mathcal T}\hat {\mathcal V}_{-k}}{12 \Pi}=\frac{i k^2
e^{4\mathcal T}\hat {\mathcal V}_{k}^+}{12 \Pi}$. The quantities
$\hat P_{\Lambda}$, $\hat \Lambda$ and  $\tau$ commutes with all
others initially. The commutator algebra could be also obtained
with the help of the Dirac brackets
\cite{Cherkas2012,Cherkas2015}.

After the definition of initial conditions for the operator
evolution (Eq. (18), see Fig. \ref{H}), the following step is to
define the Hilbert space where the quasi-Heisenberg operators act.
As we stated previously, the quasi-Heisenberg picture is an
alternative to the WDW equation, however, it turns out that for
building the Hilbert space, one should return to the Hamiltonian
(\ref{hsm}) and consider it as the WDW equation in the vicinity of
${\mathcal T}_0\rightarrow -\infty$
\cite{Cherkas2006,Cherkas2012,Cherkas2015}. Heretofore, the
momentum $P_\Lambda$ should be excluded with the help of the gauge
condition $P_\Lambda=\Pi$.

The corresponding WDW equation  in the vicinity of $\tau={\mathcal
T}_0\rightarrow -\infty$ is given as
\bea
\biggl(-i 6 \Pi\frac{\ptl }{\ptl \tau}+12
\Pi^2-\frac{1}{2}\frac{\ptl^2}{\ptl
v_0^2}~~~~~~~~~~~~~~~~~~~~~~~~~~~~~~~\nonumber\\-\sum_{k\ge
 1}\frac{\ptl}{\ptl v_k}\frac{\ptl}{\ptl
v_{k}^*}\biggr)\Psi(\tau,..v_{1}^*,v_0,v_1...)=0.~~~~~
\label{wheel}
\eea
where term $e^{4\tau }k^2 \hat v_k \hat v_{k}^+$ is omitted
because the states of the form of the wave packet will be
considered below. Let in some of this states typical value of the
square of momentum of the mode $k$ is $<\hat p_k \hat p_k^+>\sim
1/a_k$, then the typical value of $<e^{4\tau }k^2 \hat v_k \hat
v_{k}^+>\sim e^{4\tau }k^2 a_k$ due to uncertainty principle, so
it becomes negligible in the vicinity $\tau={\mathcal
T}_0\rightarrow -\infty$ which just be needed. Here
 $\hat v_k^+=v_k^*$, $\hat {p^+_k}=-i\frac{\ptl}{\ptl v^*_{k}}$.

The mean value of the quasi-Heisenberg operator
$A(\eta,\tau,v_i,\hat p_i)$ is given by  formula
\bea
  <\psi|\hat A(\eta,\tau,v_j,-i\frac{\ptl}{\ptl
v_j})|\psi>= \int\Psi^*(\tau,v_j)\hat
A(\eta,\tau,v_j,-i\frac{\ptl}{\ptl v_j})\nonumber\\
\Psi(\tau,v_j)dv_0dv_1dv_1^*\dots \biggr|_{\,\tau={\mathcal
T}_0\rightarrow -\infty }~~,~~~~~~~~~~
\label{meangen0}
\eea
where the integral over $dzdz^*\equiv\frac{\rho d\rho d\phi}{2\pi
i}$ and $z=\rho e^{i\phi}$ is understood in the holomorphic
representation \cite{Faddeev1987}. It should be noted that as well
as in the  Klein-Gordon current scalar product
\cite{Cherkas2006,Cherkas2012,Cherkas2015} there is no integration
over the variable $\tau$ in equation (\ref{meangen0}). Instead, it
is set to some quantity ${\mathcal T}_0$. For instance, in more a
general case of the equation containing the derivatives
$\frac{\ptl^2}{\ptl \tau^2}$ as well as $\frac{\ptl}{\ptl \tau}$ ,
the scalar product should contain as
 the term $i\left(\frac{\ptl\Psi}{\ptl \tau}
\Psi^*-\frac{\ptl\Psi^*}{\ptl \tau} \Psi\right)$ of the "current"
type, so the term $\Psi^*\Psi$ of the "density" type. In any case
the quantity $\tau$ should be set to some value $\mathcal T_0$
\cite{Mostafazadeh2004}. Here, the quantity ${\mathcal T}_0$  is
chosen to be initially finite, thus avoiding the singularity, but
finally the limit ${\mathcal T}_0 \rightarrow -\infty$ is taken.

 The general  solution of Eq. (\ref{wheel}) may be written in  the form of the wave
packet

\begin{widetext}
\be
  \Psi(\tau,..v_{1}^*,v_0,v_1...)=\int
C(..p_{1}^*,p_0,p_1...) \exp\left(-\frac{i}{6 \Pi}\left(12
\Pi^2+\frac{1}{2}p_0^2+\sum_{k\ge 1}p_kp_k^*\right)\tau+i
\sum_{k\ge 0} v_kp_{k}^*\right)dp_0dp_1dp_1^*...
\label{w}
\ee
\end{widetext}

In the momentum representation, the wave function (\ref{w}) takes
the form
\bea
\psi(\tau,..p_{1}^*,p_0,p_1...)=C(..p_{1}^*,p_0,p_1...)
\nonumber~~~~~~~~~~~~~\\\exp\biggl(-\frac{i}{6 \Pi}\biggl(12
\Pi^2+\frac{1}{2}p_0^2+\sum_{k\ge 1}p_kp_k^*\biggr)\tau\biggr),~~~
\label{pack}
\eea
and formula (\ref{meangen0}) for mean value looks like
\bea
  <\psi|\hat A(\eta,\tau,\mbox{i}\frac{\ptl}{\ptl
p_j},p_j)|\psi>= \int\psi^*(\tau,p_j)
~~~~~~~~~~~~~~~\nonumber\\\hat
A(\eta,\tau,\mbox{i}\frac{\ptl}{\ptl
p_j},p_j)\psi(\tau,p_j)dp_0dp_1dp_1^*\dots
\biggr|_{\,\tau={\mathcal T}_0\rightarrow -\infty}.~~~~~
\label{meangen}
\eea

For this simple model, the analytical solution exists that allows
demonstrating the calculation of mean values in detail. The
solution of Eq. (\ref{diff1}) is
\be
\tau(\eta)={\mathcal T}_0+\frac{1}{2}\ln\left(1+12 \Pi
e^{-2{\mathcal T}_0}\eta\right).
\label{tt}
\ee
First, let us consider the solution of Eq. (\ref{diff2}) in the
vicinity of $\tau\sim {\mathcal T}_0\rightarrow -\infty$. It takes
the form
\be
\hat {\mathcal V_k}(\eta)\approx \hat v_k+\frac{1}{12 \Pi} p_k^*
\ln \left(1+12 \Pi e^{-2 {\mathcal T_0} }\eta\right).
\label{field0}
\ee

If ${\mathcal T}_0$ tends to minus infinity, then the expression
(\ref{tt}) for $\tau(\eta)$ becomes
$\tau(\eta)=\frac{1}{2}\ln\left(12\Pi \eta\right)$.
 However, the expression for the operator
$\hat {\mathcal V}_k(\eta)$ diverges formally as ${\mathcal
T}_0\rightarrow -\infty$. This reflects the fact that it is
impossible to set the field values at the singularity in the
classical picture. Below we demonstrate that the quantum picture
validates the limit of ${\mathcal T}_0\rightarrow -\infty$ for the
mean observable values.

Let us consider the mean value of (\ref{field0}) over the wave
packet (\ref{pack})

\begin{widetext}
\bea
  <\psi|\hat {\mathcal V}_k|\psi>=\int (C(..p_{1}^*,p_0,p_1...))^*
\exp\left(\frac{i}{6 \Pi}(12 \Pi^2+\sum_{q\ge 0}p_qp_q^*){\mathcal
T}_0\right)\nonumber\biggl(i\frac{\ptl}{\ptl p_k}+\frac{1}{12
\Pi}p_k^* \ln \left(1+12 \Pi e^{-2 {\mathcal T_0}
}\eta\right)\biggr)\nonumber\\\exp\left(-\frac{i}{6 \Pi}(12
\Pi^2+\sum_{q\ge 0}p_qp_q^*){\mathcal
T}_0\right)C(..p_{1}^*,p_0,p_1...)dp_0dp_1dp_1^*\dots\biggr|_{\,{\mathcal
T}_0\rightarrow -\infty }\nonumber\\=\int
(C(..p_{1}^*,p_0,p_1...))^*\biggl(\frac{1}{12\Pi}p_k^*\ln(1+12\Pi
e^{-2{\mathcal T}_0}\eta)+\frac{1}{6 \Pi}p_k^*{\mathcal
T}_0+i\frac{\ptl}{\ptl
p_k}\biggr)C(..p_{1}^*,p_0,p_1...)dp_0dp_1dp_1^*\dots\biggr|_{\,{\mathcal
T}_0\rightarrow -\infty }\nonumber\\ = \int
(C(..p_{1}^*,p_0,p_1...))^*\biggl(\frac{1}{12\Pi}p_k^*\ln(12\Pi
\eta)+i\frac{\ptl}{\ptl
p_k}\biggr)C(..p_{1}^*,p_0,p_1...)dp_0dp_1dp_1^*\dots~~~~~~~
\label{calc}
\eea
\end{widetext}

One can see from Eq. (\ref{calc}) that the divergent terms with
${\mathcal T}_0\rightarrow -\infty$ cancel each other, and the
mean value of $\hat {\mathcal V}_k$ is finite. Hence, the wave
packet defined at the singularity determines the entire  evolution
of the system.

The approximate expression for $\hat {\mathcal V}_k$ has been used
above. It is valid for $\eta\sim 0$. However, it is intensional to
consider the exact expression and the contribution of  $V-$
quantum fluctuations to the $\lambda-$ evolution. The exact
solution of the equation of motion (\ref{diff2}) with $\tau(\eta)$
given by (\ref{tt}) takes the form

\begin{widetext}
\bea
  \hat {\mathcal V}_k(\eta)=\frac{\pi }{24 \Pi} \biggl( p_{k}^*
J_0\biggl(\frac{e^{2 {\mathcal T}_0} k}{12 \Pi }\biggr)
Y_0\biggl(k \biggl(\eta+\frac{e^{2 {\mathcal T}_0}}{12 \Pi
}\biggr)\biggr)-J_0\biggl(k \biggl(\eta+\frac{e^{2 {\mathcal
T}_0}}{12 \Pi
   }\biggr)\biggr) \biggl( p_{k}^* Y_0\biggl(\frac{e^{2 {\mathcal T}_0} k}{12 \Pi }\biggr)
   +k e^{2 {\mathcal T}_0} \hat v_k Y_1\biggl(\frac{e^{2 {\mathcal T}_0} k}{12 \Pi }\biggr)
   \biggr)\nonumber\\+k e^{2 {\mathcal T}_0} \hat v_k
   J_1\biggl(\frac{e^{2 {\mathcal T}_0} k}{12 \Pi }\biggr) Y_0\biggl(k \biggl(\eta
   +\frac{e^{2{\mathcal T}_0}}{12 \Pi
   }\biggr)\biggr)\biggr).~~~~
\label{solv}
\eea
\end{widetext}

\noindent Here $J_0(z), Y_0(z),  Y_1(z)$ and $ J_1(z)$ are the
Bessel functions. The second derivative of $\Lambda_0$ can be
determined from the equation of motion (\ref{diff3}), whereas its
first derivative can be determined from the Hamiltonian constraint
(\ref{hk}):
\bea
\hat \Lambda_0^\prime=\frac{1}{\tau^\prime}\left(2\tau^{\prime
2}-3\hat {\mathcal V}^{\prime 2}_0-6\sum_{k \ge 1}\hat  {\mathcal
V}^\prime_k \hat {\mathcal V}^{\prime + }_{k}+k^2 \hat {\mathcal
V}_k\hat {\mathcal V}_{k}^{+}\right).~~~~~~~
\label{firstd}
\eea
Here $\hat {\mathcal V}^{+}_{k}$ should be obtained  from Eq.
(\ref{solv}) by changing $\hat v_k\rightarrow \hat
{v^+_k}=i\frac{\ptl}{\ptl p_k^*}$, $p^*_k\rightarrow p_k$. Thus
the most intriguing  problem is the calculation mean values of
$\hat {\mathcal V}^\prime_k \hat {\mathcal V}^{\prime+ }_{k}$ and
$k^2 \hat{\mathcal V}_k\hat{\mathcal V}_{k}^{+}$, which are
constituents of Eqs. (\ref{diff3}) and (\ref{firstd}) for $\hat
\Lambda_0^{\prime}$, $\hat \Lambda_0^{\prime\prime}$. Tracing this
quantities allows calculating the $\hat \Lambda_0-$evolution.

Let us take the Gaussian form of the wave packet to determine the
evolution of the system
\be
C(..p_{1}^*,p_0,p_1...)=\prod_{k=0}^\infty N_k\exp\left(-a_k
p_kp_k^*\right),
\ee
where the constant $a_k$ determines the width of the packet for
each mode and $N_k$ is the normalization factor. The calculation
according to (\ref{meangen}) leads to the expressions  defining
the mean value of the potential energy $\Xi_k$ and the value of
the kinetic energy $K_k$ of each  mode $k\ne0$:

\begin{widetext}
\bea
  \Xi_k\equiv<\psi|k^2 \hat{\mathcal V}_k\hat {\mathcal
V}_{k}^{+ }|\psi>=\frac{k^2}{1152 a_k \Pi ^2}\biggl( \biggl(4
J_0^2(k \eta) \biggl(144 a_k^2 \Pi ^2+\log ^2\biggl(\frac{k}{24
\Pi }\biggr)
 +2 \gamma \log \biggl(\frac{k}{24 \Pi }\biggr)+\gamma
^2\biggr)\nonumber\\-4 \pi \biggl(\log \biggl(\frac{k}{24 \Pi
}\biggr)+\gamma
   \biggr) J_0(k \eta) Y_0(k \eta)+\pi ^2 Y_0^2(k \eta)\biggr)\biggr),
\nonumber\\K_k\equiv<\psi|\hat{\mathcal V}^\prime_k \hat{\mathcal
V}^{\prime +}_{k}|\psi>=\frac{k^2}{1152 a_k \Pi ^2}\biggl(
\biggl(4 J_1^2(k \eta) \biggl(144 a_k^2 \Pi ^2+\log
^2\biggl(\frac{k}{24 \Pi }\biggr)
 +2 \gamma \log \biggl(\frac{k}{24 \Pi }\biggr)+\gamma
^2\biggr)\nonumber\\-4 \pi \biggl(\log \biggl(\frac{k}{24 \Pi
}\biggr)+\gamma
   \biggr) J_1(k \eta) Y_1(k \eta)+\pi ^2 Y_1^2(k \eta)\biggr)\biggr),
\eea
\end{widetext}

where $J_0(z), Y_0(z), J_1(z)$ and $ Y_1(z)$ are the Bessel
functions and $\gamma$ is the Euler constant.

 A spatially uniform mode contains only the kinetic energy term
\[
  K_0\equiv\frac{1}{2}<\psi|{\mathcal V}^{\prime 2}_0
|\psi>=\frac{1}{1152 \,a_0^2 \Pi^2 \,\eta^2}.
\]

For further analysis, it is convenient to consider the
quasi-classical sector corresponding to late times. This insight
can be provided by expanding the Bessel function into series over
a large argument and keeping the leading terms:
\bea
  Y_0(z)\approx\frac{1}{\sqrt{\pi z}}\Biggl(\left(-\frac{9}{128
z^2}-\frac{1}{8 z}+1\right) \sin
(z)~~~~~\nonumber\\+\left(\frac{9}{128
z^2}-\frac{1}{8 z}-{1}\right) \cos (z)\Biggr),\nonumber\\
  J_0(z)\approx\frac{1}{\sqrt{\pi z}}\Biggl(\left(-\frac{9}{128
z^2}+\frac{1}{8   z}+{1}\right) \sin
(z)~~~~~\nonumber\\+\left(-\frac{9}{128
z^2}-\frac{1}{8  z}+{1}\right) \cos (z)\Biggr),\nonumber\\
  J_1(z)\approx\frac{1}{\sqrt{\pi z}}\Biggl(\left(\frac{15}{128
z^2}+\frac{3}{8  z}+{1}\right) \sin
(z)~~~~~\nonumber\\+\left(-\frac{15}{128
z^2}+\frac{3}{8  z}-{1}\right) \cos (z)\Biggr),\nonumber\\
  Y_1(z)\approx\frac{1}{\sqrt{\pi z}}\Biggl(\left(-\frac{15}{128
  z^2}+\frac{3}{8   z}-{1}\right) \sin
(z)~~~~~\nonumber\\+\left(-\frac{15}{128  z^2}-\frac{3}{8
z}-{1}\right) \cos (z)\Biggr)\nonumber.
\eea

Then, a simple estimation results from replacement  the
oscillating multipliers by their time-averaged values as $
\cos^2(k \eta)\rightarrow \frac{k}{2\pi}\int_0^{2\pi/k}\cos^2(k
\eta)d \eta=\frac{1}{2}$, $\sin^2(k \eta)\rightarrow \frac{1}{2}$,
and $\sin(k \eta)\cos(k \eta)\rightarrow 0$.

Using Eqs. (\ref{diff3}) and (\ref{firstd}) we get

\begin{widetext}
\bea
  <\psi|\hat \Lambda_0^{\prime}|\psi>\approx
\frac{1}{\eta}\left(1-\frac{1}{96 {a_0} \Pi ^2}\right)-\sum_{k\ge
1}\frac{k F_k}{6 \Pi^2\pi a_k}+\frac{12 a_k k}{\pi }
+\frac{1}{\eta^2}\left(\frac{F_k}{48\pi\Pi^2 k a_k}+\frac{3 a_k}{2
\pi k}\right),
   \label{first}
\eea
\bea
  <\psi|\hat
\Lambda_0^{\prime\prime}|\psi>\approx-\frac{1}{\eta^2}\left(1-\frac{1}{96
{a_0} \Pi ^2}\right)+\frac{1}{\eta^3}\sum_{k\ge
1}\frac{F_k}{24\pi\Pi^2 k a_k} +\frac{3 a_k}{\pi  k},
    \label{second}
\eea
\end{widetext}

where $F_k=\biggl(\frac{\pi^2 }{8}+\frac{\gamma
^2}{2}+\frac{1}{2}\ln ^2\left(\frac{k}{24 \Pi }\right)+\gamma
   \ln \left(\frac{k}{24 \Pi }\right)\biggr)$.

 It should be noted that Eq. (\ref{second})
 describing the averaged second
derivative of $\Lambda_0$ in a sense of  the time-averaged
 evolution can be obtained from Eq. (\ref{first}) by
the differentiation over $\eta$. Turning to a continuous limit of
$\sum_k \rightarrow \frac{1}{2\pi}\int dk $, we can see that the
second term in Eq. (\ref{first}), corresponding to the vacuum
energy, diverges for any asymptotic of $a_k$ at large $k$.

The most divergent term $\frac{k F_k}{6 \Pi^2\pi a_k}+\frac{12 a_k
k}{\pi }$ vanishes under differentiation of Eq. (\ref{first}). The
remained  term is the mean value of the difference of the
potential and kinetic energies of field oscillators, and has been
considered in Ref. \cite{Cherkas2007} for the Friedman universe.
It has been found that this term defines the value of the
acceleration parameter of universe, which is compatible with the
observed one. One has note, that the UV cut-off of momenta was
used for the estimates \cite{Cherkas2007} for the Friedman
universe. The present-day universe  expands isotropically, so one
cannot compare the results of the above calculations with some
observational values directly. The early stages of the universe
could be highly anisotropic \cite{Belinsky1970}. Particle creation
during the anisotropic cosmological expansion and its back
reaction to the metric have been considered \cite{Lukash1974}. It
is interesting that the authors of Ref. \cite{Lukash1974}  faced
the necessity to set initial conditions for the evolution. They
were forced to begin the evolution from a certain artificial
moment of time. As we have seen above in the quasi-Heisenberg
picture there exists fundamental possibility to set the initial
conditions at the singularity itself and therefore to improve the
analysis of Ref. \cite{Lukash1974}.

\section{Evolution determined by the vacuum state}\label{sect3}

In the considered gauge  the background variable $\tau$ is not
quantum. For this particular case, one can use the ordinary
quantization using the creation and annihilation operators. Thus,
we consider the quantization of the field $V$ against the
time-dependent background $\tau(\eta)=\frac{1}{2}\ln(12 \Pi
\eta)$. In this case, the field $V$ is represented as
\cite{Birrell1982}
\be
{\mathcal V}_k(\eta)=\sum_k \hat{\mbox a}_k u_k(\eta)+\hat{\mbox
a}_k^+ u_k^*(\eta),
\ee
where $ [{\hat {\mbox{a}}_k},{\hat {\mbox{a}}}_k^+]=1. $

\noindent The function $u_k(\eta)$ should satisfy the condition
\be
e^{2 \tau (\eta)} \left({u_k^*}(\eta) u_k^\prime(\eta)-u_k(\eta)
u_k^{*\prime}(\eta)\right)=i.
\label{rel0}
\ee

The mean values of the kinetic and potential energies of the mode
$k$ in a vacuum state  equal to
\bea
\Xi_k= <0|k^2 {\mathcal V}_k{\mathcal V}_{k}^{+
}|0>=k^2 u^{*}_k u_k,\nonumber \\
K_k=<0|{\mathcal V}^\prime_k {\mathcal V}^{+\prime
}_{k}|0>=u^{*\prime}_k u^{\prime}_k.
\eea
Thus, one has to determine the functions $u_k$. The vacuum state
is defined as a state vanishing under the action of the
annihilation operator: $\hat {\mbox{a}}_k|0>=0$. However, the
definition of $u_k$ is ambiguous. It should be noted that there
exists a family of functions $u_k$ which satisfy Eq. (\ref{rel0})
and are interrelated by the Bogolubov's transformation. It was
shown \cite{Anishchenko2009} the vacuum state could be defined
through the minimization of some functional containing the
difference of the potential and kinematic energies of field
oscillators. In such a way one comes to the function
\be
u_k(\eta)=\frac{1}{4} \sqrt{\frac{\pi }{3\Pi}}\,
H_0^{(2)}(|k|\eta),
\ee
where $H_0^{(2)}(z)$ is the Hankel function of the second kind.
There is no particle (i.e., graviton) creation here,  because the
difference of the kinetic and potential energies is not an
oscillating quantity \cite{Anishchenko2009}.

  Using the asymptotics of the Hankel
function for large arguments,
\[
 H_0^{(2)}(z)\approx\sqrt{\frac{2}{\pi z}}\,e^{-i(z-\pi/4)}\left(1+\frac{i}{8 z}-\frac{9}{128 z^2}\right)
\]
 one can obtain for the mean values of $\hat\Lambda_0^\prime$ and $\hat\Lambda_0^{\prime\prime}$ over vacuum state
\bea
<0|\hat\Lambda_0^\prime|0>\approx\frac{1}{\eta}-\sum_{k\ge
1}\frac{k}{\Pi}+\frac{1}{8 k \eta^2
\Pi},\nonumber\\
<0|\hat\Lambda_0^{\prime\prime}|0>\approx
-\frac{1}{\eta^2}+\sum_{k\ge 1}\frac{1}{4 k\eta^3 \Pi}.
\label{vv}
\eea
It is interesting to compare the above results with those from the
quasi-Heisenberg quantization. For this aim one has to find the
value $a_k$ in Eqs. (\ref{first}),(\ref{second}) which minimizes
the constant part contribution $\frac{k F_k}{6 \Pi^2\pi
a_k}+\frac{12 a_k k}{\pi }$ of every mode to $\Lambda_0^\prime$
given by Eq. (\ref{first}). That gives
$a_k=\frac{1}{6\Pi}\sqrt{\frac{F_k}{2}}$. Substitution of this
value into Eqs. (\ref{first}) and (\ref{second}) leads to
\bea
  <\psi|\hat \Lambda_0^{\prime}|\psi>\approx
\frac{1}{\eta}\left(1-\frac{1}{96 {a_0} \Pi
^2}\right)~~~~~~~~~~~~~~~~~~~~~~~~~~~~~\nonumber\\-\sum_{k\ge
1}\sqrt{1+\frac{4}{\pi^2}\left(\gamma+\ln\left(\frac{k}{24
\Pi}\right)\right)^2}\left(\frac{k}{\Pi} +\frac{1}{8 k \eta^2
\Pi}\right),~~~~~
   \label{first1}
\eea
\bea
  <\psi|\hat
\Lambda_0^{\prime\prime}|\psi>\approx-\frac{1}{\eta^2}\left(1-\frac{1}{96
{a_0} \Pi ^2}\right)~~~~~~~~~~~~~~~~~~~~~\nonumber\\+\sum_{k\ge
1}\sqrt{1+\frac{4}{\pi^2}\left(\gamma+\ln\left(\frac{k}{24
\Pi}\right)\right)^2}{\frac{1}{4 k\eta^3 \Pi}}.~~~~~
    \label{second1}
    \eea
The comparison  with Eq. (\ref{vv}) demonstrates that the
non-vanishing term supplements a vacuum state  term in the
quasi-Heisenberg quantization scheme.

Thus, any momentum wave packet defined at singularity gives an
inevitable counterpart corresponding to a matter (in this model
"matter" consists of gravitational wave quants). There is no need
in ``matter creation from nothing'' in the quasi-Heisenberg
picture, because it exists primordially.

Let us briefly discuss the vacuum energy and its relation to
singularity. Before regularization, the expressions for the mean
values of  $\Lambda_0^\prime$ and $\Lambda_0^{\prime\prime}$ are
singular.  Regularization  of the influence of quantized
gravitational waves to a background have been considered
\cite{Berger1982,Hussein1987}. The author of Ref.
\cite{Berger1982} has found that the singularity disappears that
occurs because the substraction, that she uses in a
regularization, affects the classical terms. However, the author
of Ref. \cite{Hussein1987} stated that the singularity still
remains. His argumentation is that for coherent states the mean
values in classical and quantum pictures must coincide. For this
purpose  he took an appropriate ordering of the creation and
annihilation operators in calculating the mean values. However, it
should be noted that the vacuum state is a particular case of the
coherent state. Thus, it is not surprising that the vacuum
fluctuations do not contribute to evolution (i.e. do not affect
the singularity) according to \cite{Hussein1987}.

In the previous section it has been conjectured that a difference
of the potential and kinetic energies has a physical meaning if
one uses the UV cut-off. It comes from the fact that difference of
the potential and kinetic energies of field oscillators gives a
value of the universe acceleration compatible with observations
\cite{Cherkas2007}. Thus, it seems that only the main divergence
(also existing in the Minkowsky space-time) should be subtracted.

\section{Outlook}

As was discussed in the previous section, we cannot say infallibly
whether singularity exists or not without a fundamental theory of
regularization of the vacuum energy. However, earlier it have been
found no vacuum energy problem in the toy two dimensional model
considering string on the curved background \cite{Cherkas2012},
because the cosmological expansion is simply a motion of the
string center of mass. Fluctuations, including vacuum ones, do not
affect the motion of the string center of mass, i.e. the
cosmological expansion. Mathematically, this looks as a
compensation of scale factor fluctuations by fluctuations of the
matter fields \cite{Cherkas2012}.

On the other hand, in GR there exists the Isaacson theorem
\cite{Isaacson1968} which states that evolution in the mean is
determined by the energy-momentum tensor of excitations
(perturbation). Thus, in the theories for which  the Isaacson
theorem is valid the vacuum energy problem emerges. Roughly, since
the Isaacson theorem does not differ the vacuum fluctuations from
the excitations under vacuum, the vacuum fluctuations contribute
to the mean evolution.

Being capable of the solving the vacuum energy problem the
theories where the Issacson theorem does not exist, are  beyond
the GR frameworks. One may assume, that a quantum version of the
Isaacson theorem should be developed for GR to differ vacuum and
non-vacuum fluctuations. Also, it seems important to investigate
the connection of the Isaacson theorem with the conformal
invariance of the gravity theories \footnote{Recent interesting
example of the conformally invariant theory of gravity have been
developed \cite{Gomes2011,Smolin2014}.}.

 To summarize, as it was
shown in section \ref{sect2}, it is possible to describe the
universe evolution before regularization by a wave packet
definition at singularity regardless a regularization procedure.
It should be emphasized that the wave packet determined at the
singularity is not only an ``informational seed'' but it is also
responsible for the part of the matter in the universe because the
gravitons (and, in the general case, the quants of matter fields)
appears inevitably at the late time evolution.

 It
would be interesting to consider a quantum picture of the general
3+1 BKL-solution \cite{Belinsky1970} in the framework of the
quasi-Heisenberg picture including building of the corresponding
wave packet at singularity. This work is in progress
\cite{Cherkas2014}.




\end{document}